\documentstyle[aps,prl,epsf,twocolumn,floats,graphicx]{revtex}

\setkeys{Gin}{width=0.9\linewidth}

\newcommand\yupd{Y$_{\rm 1-x}$U$_{\rm x}$Pd$_3$}
\newcommand\ucu{UCu$_{\rm 5-x}$M$_{\rm x}$}
\newcommand\uthpd{Th$_{\rm 1-x}$U$_{\rm x}$Pd$_{2}$Al$_{3}$}
\begin{document}
\draft
\title{\boldmath Evidence for a common physical description of
non-Fermi-liquid behavior in $f$-electron systems}
\author{M.~C.~de~Andrade, R.~Chau, R.~P.~Dickey, N.~R.~Dilley,
E.~J.~Freeman, D.~A.~Gajewski, and M.~B.~Maple}
\address{Department of Physics and Institute for Pure and Applied
Physical Sciences,\\
University of California, San Diego, La Jolla, CA 92093}
\author{R.~Movshovich}
\address{Los Alamos National Laboratory, Los Alamos, NM 87545}
\author{A.~H.~Castro~Neto and G.~E.~Castilla}
\address{Department of Physics, University of California, Riverside,
CA 92521}
\author{B. A. Jones}
\address{IBM Almaden Research Center, San Jose, CA 95120-6099}
\date\today
\maketitle

\begin{abstract}

The non-Fermi-liquid (NFL) behavior observed in the low
temperature
specific heat $C(T)$ and magnetic susceptibility $\chi(T)$ of
{\it f}-electron systems is analyzed within the context of a recently
developed theory based on Griffiths singularities.
Measurements of $C(T)$ and $\chi(T)$ in the systems
Th$_{\rm 1-x}$U$_{\rm x}$Pd$_{2}$Al$_{3}$,
Y$_{\rm 1-x}$U$_{\rm x}$Pd$_3$,
and UCu$_{\rm 5-x}$M$_{\rm x}$ (M = Pd, Pt) are found
to be consistent with
$C(T)/T~\propto~\chi(T)~\propto~T^{-1+\lambda}$
predicted by this model with $\lambda <1$ in the NFL regime.
These results suggest that the NFL properties observed in a wide variety
of {\it f}-electron systems can be described
within the
context of a common physical picture.

\end{abstract}

Transport, thermal, and magnetic measurements on a number of
chemically
substituted rare earth and actinide compounds have revealed low
temperature physical properties that show striking departures from the
predictions of Fermi-liquid theory\cite{ITP}.
Several theoretical models have been developed to account for the
non-Fermi-liquid (NFL) behavior observed in {\it f}-electron materials.
These models include a multichannel Kondo effect of magnetic or electric
origin\cite{NOZIERES80,SCHLOTTMAN93,COX87},
fluctuations of an order parameter in the vicinity of a second
order phase transition at  $T = 0$~K
\cite{MILLIS93,TSVELIK93,SACHDEV95},
a disordered distribution of Kondo temperatures
\cite{BERNAL95,MIRANDA96}, and an electron polaron model
for heavy fermion systems \cite{LIU97}.
However, none of these models has been able to account for all of the NFL
characteristics observed in the wide variety of systems that belong to
this new class of strongly correlated {\it f}-electron materials.
Three of us (A.~H.~C.~N., G.~E.~C., and B.~A.~J.) have recently proposed a
model
where NFL behavior is associated with the proximity to a quantum critical
point and the formation of magnetic clusters in the paramagnetic phase
due to the competition between the Kondo effect
and the Ruderman-Kittel-Kasuya-Yosida (RKKY) interaction in the presence
of  magnetic anisotropy and disorder inherent in alloyed materials
\cite{CASTRONETO98A}.
\begin{table*}
	\centering
	\caption{Exponent $\lambda$ obtained from fits of Eq. (1) to
	specific heat
	 ($\lambda_{C}$) and magnetic susceptibility ($\lambda_{\chi}$)
	 data for the {\it f}-electron systems shown in Figs. 1 and 2.}
\vskip 1.5ex
	\begin{tabular}{ccccccccccc}
	& \multicolumn{6}{c}{Th$_{{\rm 1-x}}$U$_{{\rm%
x}}$Pd$_{2}$Al$_{3}$}
&\multicolumn{2}{c}{UCu$_{{\rm 5-x}}$Pd$_{{\rm x}}$}  \\
            		\cline{2-7}\cline{8-9}
			x  & 0  & 0.2 & 0.4 & 0.6 & 0.8 & 0.9 & 1.0 & 1.5
& \raisebox{1.5ex}[0pt][0pt]{UCu$_{4}$Pt} & \raisebox{1.5ex}[0pt][0pt]
			{Y$_{0.8}$U$_{0.2}$Pd$_{3}$} \\
			\hline
			$\lambda_{C}$ & 1 & 0.85 & 0.81 & 0.84 & 0.81 & 1
			& 0.72 &
			0.81 & 0.83 & 0.76  \\
			$\lambda_{\chi}$  & 0.87 & 0.6 & 0.63 & 0.63 & 0.6
			& --- & 0.72
			& 0.78 & 0.77 & 0.70\\
		\end{tabular}
	\label{tbl:1}
\end{table*}
This model predicts that various physical properties diverge with
decreasing
temperature as weak power laws of temperature and that this behavior
persists over appreciable ranges of substituent concentration, similar to
what has been observed in a number of {\it f}-electron materials.

In this letter, we compare low temperature specific heat $C(T)$ and
magnetic
susceptibility $\chi(T)$ data for the ordered {\it f}-electron sublattice
systems \ucu\ (M = Pd, Pt) (Ref. \cite{ANDRAKA93,CHAU96}) and the
disordered
{\it f}-electron sublattice systems \uthpd\ (Ref. \cite{MAPLE95A})
and \yupd\ (Ref. \cite{MAPLE95B}) with the predictions of the new model
proposed in Ref. \cite{CASTRONETO98A}.
In particular, we compare fits of our specific heat data with logarithmic
functions of temperature, as predicted in several previously developed
theories of NFL behavior, to fits with power laws in temperature as
suggested by this new model.
The data are all derived from our own measurements, some of which
are new and reported here for the first time and others of which
have been published previously
\cite{CHAU96,MAPLE95A,MAPLE95B,SEAMAN91,MAPLE94}.
We find that the NFL behavior of $C(T)$ and $\chi (T)$ in these compounds
is consistent with the predictions of the model proposed in
Ref. \cite{CASTRONETO98A}.
Specifically, $C(T)/T$ and $\chi (T)$ can be described by divergent power
laws in temperature at the lowest temperatures for all of the compounds
investigated.
Our results suggest that this general scenario holds promise for
understanding the NFL behavior in this new class of {\it f}-electron
materials.

Details of the procedures used to prepare the polycrystalline uranium
compounds studied in this work are described
elsewhere\cite{CHAU96,MAPLE95B}.
Magnetization $M(T)$ measurements were performed after field cooling a
sample to the lowest temperature using a commercial SQUID magnetometer
(Quantum Design) in fields of $0.5$ and $1$ tesla and at temperatures in
the range $1.8~{\rm K} \leq T \leq 300$~K.
The low temperature $M(T)$ data ($0.4~{\rm K} \leq T \leq 2$~K)
were acquired with a $^3$He Faraday magnetometer (FM).
Heat capacity measurements were made in a $^3$He semi-adiabatic
calorimeter with a standard heat-pulse technique.

\begin{figure}[tbp]
	\centering
	\includegraphics{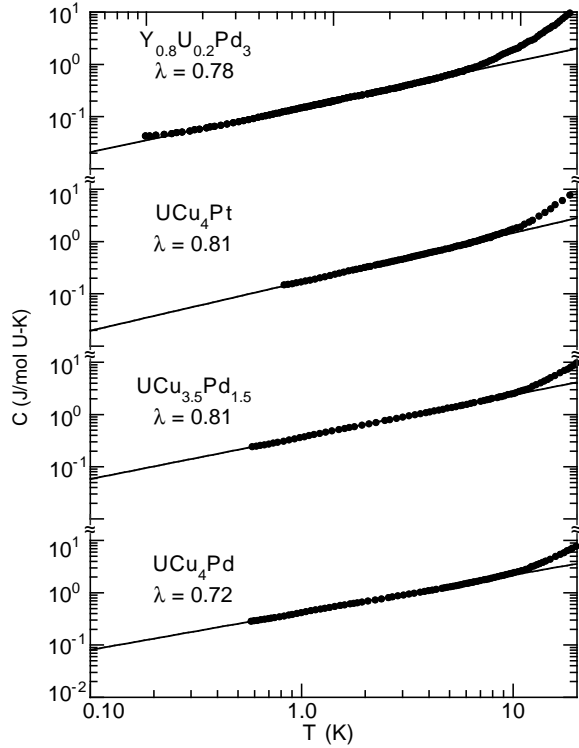}
	\caption{Log-log plot of specific heat $C$ vs temperature $T$ for
several 
	U-based NFL systems.  Solid lines are fits of the data to Eq. 1.}
	\label{fig:1}
\end{figure}

\begin{figure}[tbp]
	\centering
	\includegraphics{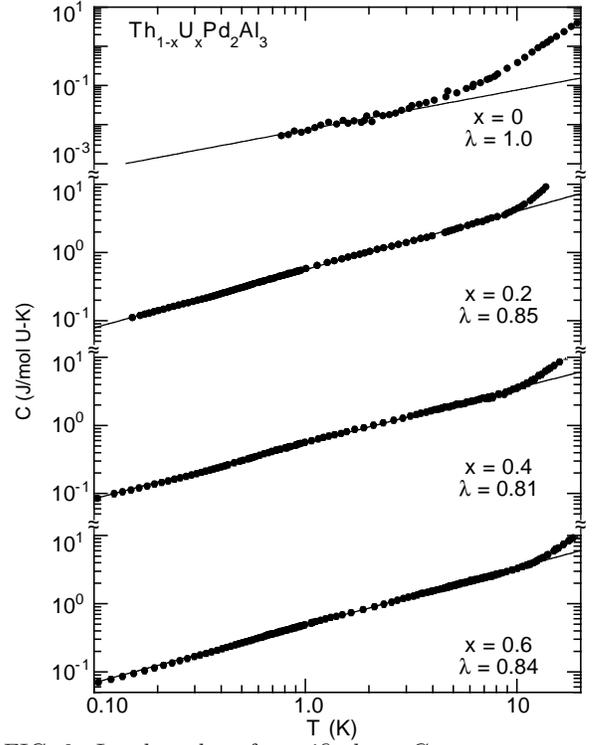}
	\caption{Log-log plot of specific heat $C$ vs temperature $T$
	for the \uthpd\ system for several values of x. Solid lines
	are fits of the data to Eq. 1.}
	\label{fig:2}
\end{figure}

Log-log plots of $C$ vs $T$ in the NFL regime are shown in Fig.\ref{fig:1}
for several samples in the \yupd\ (${\rm x} = 0.2$) and \ucu\
(M = Pt, ${\rm x} = 1$; M = Pd, ${\rm x} = 1, 1.5$) systems and
in Fig.\ref{fig:2} for the \uthpd\ (${\rm x} = 0, 0.2, 0.4, 0.6$) system.
In these figures, the error bars are smaller than the size
of the symbols.
The solid lines represent least squares fits of the expression 
relating the specific heat $C$ to the magnetic susceptibility $\chi$,
given by
\begin{equation}
		C(T)/T  \propto \chi(T)  \propto T^{-1+\lambda}
\label{Eq1}
\end{equation}
at low temperatures, where $\lambda$ is a parameter determined
by the best fit.
The values of $\lambda$ for different compounds and/or different
compositions x of the chemical substituent are indicated in
Figs.\ref{fig:1} and \ref{fig:2} and are given in Table 1.
To test the quality of the fits, the reduced chi square,
$\chi^{2}_\nu$,
was calculated\cite{BEVINGTON92} and will be discussed below.
The log-log plots of $C$ vs $T$ (Figs.\ref{fig:1} and \ref{fig:2})
reveal that a power law
with $\lambda < 1$ provides an excellent description of the data for all
the curves.
For the \uthpd\  system, the power law describes the data  from $0.1$~K
up to $14$~K.
In the \yupd\ and \ucu\ systems, the best fit was achieved
for $0.4~{\rm K} \leq T \leq 5$~K.
Notice that for ThPd$_{2}$Al$_{3}$ we get $\lambda=1$ as expected for a
Fermi
liquid.

\begin{figure}[tbp]
	\centering
	\includegraphics{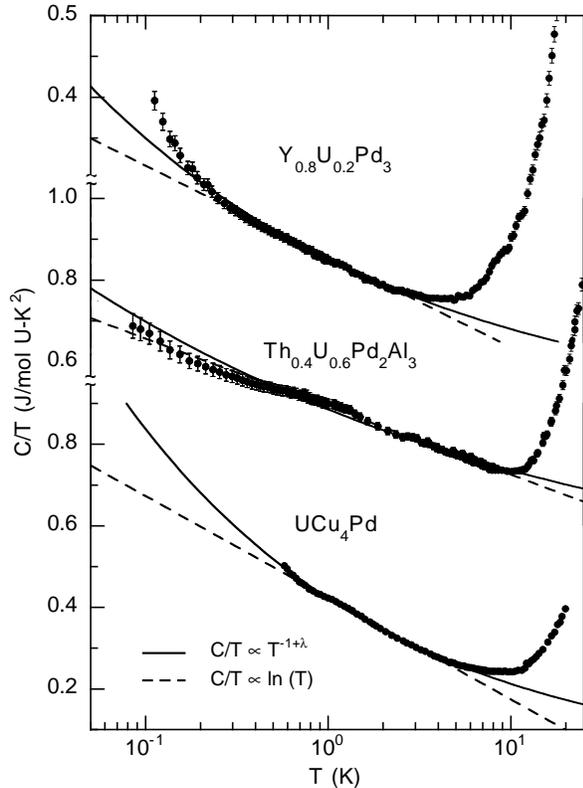}
	\caption{Semilog plot of specific heat $C$ vs temperature $T$ for 
	selected U-based NFL compounds. Solid and dashed lines are fits to
	the expressions indicated in the figure.}
	\label{fig:3}
\end{figure}

In order to provide a more direct comparison between power
law and logarithmic behavior, plots of $C/T$ vs $\log T$ are shown in
Fig.\ref{fig:3} for three selected compounds, UCu$_{4}$Pd,
Th$_{0.4}$U$_{0.6}$Pd$_{2}$Al$_{3}$, and
Y$_{0.8}$U$_{0.2}$Pd$_{3}$.
The upturns in $C(T)/T$ at high temperatures are due to the 
phonon contributions and Schottky anomalies arising from crystalline 
electric field splitting of the U $5{\it f}$ Hund's rule
ground state multiplet and are excluded from the fitting range.
These contributions were not subtracted because they could not be 
estimated with sufficient accuracy.
In this figure, both power law (solid lines) and logarithmic
(dashed lines) temperature dependences were fit to the data over the same
temperature range, and $\chi^{2}_\nu$ was then calculated to assess the
quality of each fit.
\begin{table}[b]
	\centering
	\caption{Calculated values of the reduced chi square
	$\chi^{2}_\nu$ from fits to specific heat data shown in Fig. 3.}
\vskip 1.5ex
	\begin{tabular} {lccc}
		& {Th$_{0.4}$U$_{\rm 0.6}$Pd$_{2}$Al$_{3}$}
		& {UCu$_{4}$Pd}
		& {Y$_{0.8}$U$_{0.2}$Pd$_{3}$} \\
		\hline
		   $C/T \propto \ln T$ & 0.31 & 1.57 & 3.48 \\
		 $C/T \propto T^{-1+\lambda}$ & 0.94 & 1.03 & 0.97 \\
	\end{tabular}
	\label{tbl:2}
\end{table}
Table 2 shows the calculated values of $\chi^{2}_\nu$ for each of the
compounds in Fig.3.
Recall that for an optimum fit, $\chi^{2}_\nu$ is close
to $1$\cite{BEVINGTON92}.
Although both the logarithmic and power law descriptions of the specific
heat agree with the experimental data within the experimental resolution,
the reduced chi square indicates that a power law provides a better
description of the data than a logarithmic relation over the same
temperature range.

\begin{figure}[tbp]
	\centering
	\includegraphics{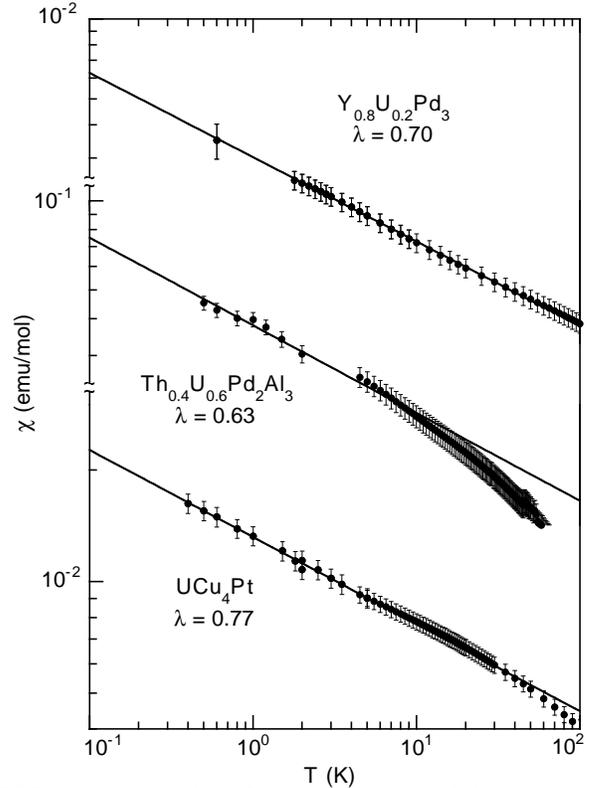}
	\caption{Log-log plot of magnetic susceptibility $\chi$ vs
	temperature $T$ for selected U-based NFL compounds. Solid lines
	are fits of the data to Eq. 1.}
	\label{fig:4}
\end{figure}

Plots of the magnetic susceptibility $\chi(T)$ vs $\ln T$ for
UCu$_{4}$Pt, Th$_{0.4}$U$_{0.6}$Pd$_{2}$Al$_{3}$ and
Y$_{0.8}$U$_{0.2}$Pd$_3$
are shown in Fig. \ref{fig:4}.
The data have been fit with power laws (Eq.\ref{Eq1}) and give values
of $\lambda$ that are close to the values obtained from the power law
fits to the $C(T)$ data.
Values of $\lambda$ from these fits are given in Table 1.
The magnetic susceptibility was obtained from the linear part of the
magnetization $M$ vs magnetic field $H$ isotherm at low fields
$\mu H/kT \ll 1$, where $\mu$ is the magnetic moment and $k$ is
Boltmann's constant.
This procedure was especially important at low temperatures where the
$M$ vs $H$ curves display negative curvature which is more pronounced
at lower temperatures.
Assuming the negative curvature of $M(H)$ is intrinsic, this method of
extracting $\chi(T)$ suggests that the origin of the NFL behavior is
magnetic in nature, since $C(T)/T$ and $\chi (T)$ have the same
temperature dependence.
We also note that the values of $\lambda$ obtained from magnetization
measurements on the \uthpd\ system reflect the average of the
magnetization over crystallites with a hexagonal structure oriented in
different directions, which may explain the difference between the
values of $\lambda$ from specific heat and susceptibility in this 
system.
In a single crystal, we expect better agreement
between the specific heat and the magnetization \cite{NAKOTTE96}.

We have shown from our experimental data that the NFL
behavior observed in $C(T)$ and $\chi (T)$ given in Eq.~\ref{Eq1}
is consistent with the existence of Griffiths singularities.
The values of $\lambda$ obtained from the $C(T)$ and $\chi(T)$ data agree
with one another within experimental resolution, suggesting that the NFL
behavior originates from magnetic interactions. The U-based systems
investigated in this work have all of the prerequisites
for the Griffiths phase scenario: the Kondo effect, RKKY interactions
between the U magnetic moments, magnetic anisotropy due to spin-orbit
interactions, and disorder associated with the chemical
substitutions.
Furthermore, the imaginary part of the frequency dependent susceptibility
$\chi''(\omega)$ of UCu$_{3.5}$Pd$_{1.5}$ and UCu$_{4}$Pd, derived from
neutron scattering measurements on these materials\cite{ARONSON95},
is described well by the Griffiths phase result
$\chi''(\omega) \propto \omega^{-1+\lambda} \tanh(\omega/T)$
\cite{CASTRONETO98A}
with a value $\lambda \approx 0.7$, in good agreement with the values of
$\lambda$ determined from the above analysis of the $C(T)$ and
$\chi (T)$ measurements.
NMR and $\mu$SR experiments\cite{BERNAL95} on these same compounds reveal
a distribution of
susceptibilities given by $\delta\chi/\chi \propto T^{-\lambda/2}$ which
is also consistent with the presence of a Griffiths phase at low
temperatures\cite{CASTRONETO98A}.
Finally, recent NMR and $\mu$SR experiments on UCu$_{\rm 5-x}$Pd$_{\rm x}$
\cite{BERNAL95}
and pressure experiments on the NFL system CeRh$_{2}$Si$_{2}$ and
CeRh$_{\rm 2-x}$Ru$_{\rm x}$Si$_{2}$\cite{GRAF97} indicate that disorder
plays
an especially important role in NFL behavior.
In addition, the relatively large range of substituent concentration x
over
which the NFL behavior extends in certain NFL systems is consistent with
the notion of a Griffiths phase since these systems have rather large
values of $T_{\rm K}$ ($\lesssim 10^{2}$~K) and the range of x over which
the Griffiths phase extends is predicted to be larger for larger values
of $T_{\rm K}$.
Thus, there is sufficient evidence to suggest that the Griffiths phase
model is a viable candidate for the NFL properties of the {\it f}-electron
systems investigated herein.
Moreover, recent calculations yield a linear temperature dependence of
the resistivity in these systems due to electron scattering by magnetic
clusters \cite{CASTRONETO98B} in agreement with transport data \cite{ITP}.

We point out that in some of these systems it is also
possible that a quadrupolar Kondo effect, due to the exchange 
interaction between the
quadrupolar moment of a $\Gamma_3$ non-magnetic doublet ground
state of U and the conduction electrons, may occur and lead
to NFL behavior \cite{SEAMAN91}.
However it is not clear that the multichannel Kondo and Kondo disorder
models,
which are single ion models, are applicable at the relatively large
substituent concentrations where NFL behavior is often observed in
{\it f}-electron materials.
Models based on the existence of quantum critical points are known to
produce power
laws in thermodynamic properties\cite{MILLIS93}.
The NFL behavior can be interpreted as a generic feature in the vicinity
of a quantum critical point\cite{SACHDEV95}.
However, in the absence of disorder and at zero temperature, the quantum
critical regions reduce to points in the phase diagram and are not general
enough to explain the broad range of substituent concentrations over
which NFL behavior is observed.
While each of these models have varying degrees of success in describing
the temperature dependence of the physical properties in these materials,
none has the general applicability to a wide range of {\it f}-electron
systems, nor the apparent internal consistency
found in our analysis in terms of Griffiths singularities.

We acknowledge illuminating converstations with M.~C.~Aronson,
W.~P.~Beyermann, D.~L.~Cox, D.~E.~MacLaughlin, R.~A.~Robinson,
and J.~D.~Thompson.
Research at UCSD was supported by the National Science Foundation under
Grant No.  DMR-97-05454 and by the UC CLC LA 95-0519-BM.
In addition, equipment used in the research at UCSD was provided by the
National Science Foundation under Grant No. DMR-94-03836.
A.~H.~C.~N. acknowledges support from the Alfred P. Sloan foundation.


\begin{references}

\bibitem{ITP} See, for instance, {\it Proceedings of the Conference on
Non-Fermi Liquid Behavior in Metals, Santa Barbara, 1996}, edited by P.
Coleman, M. B. Maple, and A. Millis, J. Phys.: Condens. Matter {\bf 8},
(1996).

\bibitem{NOZIERES80}P.~Nozi\`eres and A.~Blandin, J.~Phys.~(France)
{\bf 41}, 193 (1980).

\bibitem{SCHLOTTMAN93}P.~Schlottman and P.~D.~Sacramento, Adv.~Phys.
{\bf 42}, 641 (1993).

\bibitem{COX87}D.~L.~Cox, Phys.~Rev.~Lett. {\bf 59}, 1240 (1987).

\bibitem{MILLIS93}A.~J.~Millis, Phys.~Rev.~B {\bf 48}, 7183 (1993).

\bibitem{TSVELIK93}A.~M.~Tsvelik and M.~Reizer, Phys.~Rev.~B {\bf 48},
9887 (1993).

\bibitem{SACHDEV95}S. Sachdev, N. Read, and R. Oppermann, Phys. Rev. B 
{\bf 52}, 10286 (1995).

\bibitem{BERNAL95} O. O. Bernal, D. E. MacLaughlin, H. G. Lukefahr, and
B. Andraka, Phys. Rev. Lett. {\bf 75}, 2023 (1995).

\bibitem{MIRANDA96} E. Miranda, V. Dobrosavljevi\'c, and G. Kotliar,
J. Phys.: Condens. Matter {\bf 8}, 9871 (1996).

\bibitem{LIU97}S.~H.~Liu, Physica B {\bf 240}, 49 (1997).

\bibitem{CASTRONETO98A}A. H. Castro Neto, G. Castilla, and B. A. Jones,
cond-mat/9710123 (unpublished).

\bibitem{ANDRAKA93}B. Andraka and G. R. Stewart, Phys. Rev. B
{\bf 47}, 3208 (1993). 

\bibitem{CHAU96}R.~Chau and M.~B.~Maple, J.~Phys.:~Condens. Matter
{\bf 8}, 9939 (1996).

\bibitem{MAPLE95A}M.~B.~Maple  {\it et al.}, J.~Low~Temp.~Phys. {\bf 99},
223
(1995).

\bibitem{MAPLE95B}M.~B.~Maple {\it et al.}, J.~Phys.~Chem.~Solids {\bf
56},
1963 (1995).

\bibitem{SEAMAN91}C.~L.~Seaman {\it et al.}, Phys.~Rev.~Lett. {\bf 67},
2882
(1991).

\bibitem{MAPLE94}M.~B.~Maple {\it et al.}, J.~Low~Temp.~Phys. {\bf 95},
225
(1994).

\bibitem{BEVINGTON92}P.~R.~Bevington and D.~K.~Robinson, {\it Data
Reduction and Error Analysis for the Physical Sciences} (McGraw-Hill, New
York, 1992).

\bibitem{NAKOTTE96}H.~Nakotte {\it et al.}, Phys.~Rev.~B {\bf 54}, 12176
(1996).

\bibitem{ARONSON95}M.~C.~Aronson {\it et al.}, Phys.~Rev.~Lett. {\bf 75},
725
(1995).

\bibitem{GRAF97}T.~Graf {\it et al.}, Phys.~Rev.~Lett. {\bf 78}, 3769
(1997).




\bibitem{CASTRONETO98B}A.~H.~Castro Neto {\it et al.}, (unpublished).

\end{references}
\end{document}